\documentclass[journal]{IEEEtran}
\ifCLASSINFOpdf
  \usepackage[pdftex]{graphicx}
  \graphicspath{{../pdf/}{../jpeg/}}
  \DeclareGraphicsExtensions{.pdf,.jpeg,.png}
\else
  \usepackage[dvips]{graphicx}
  \graphicspath{{../eps/}}
  \DeclareGraphicsExtensions{.eps}
\fi

\usepackage{graphicx}
\usepackage{graphics}
\usepackage{epsfig}\usepackage{epstopdf}
\usepackage{epstopdf}
\usepackage{stfloats}
\usepackage[cmex10]{amsmath}
\usepackage{algorithmic}
\usepackage{array}
\usepackage{mdwmath}
\usepackage{mdwtab}
\usepackage{graphicx}
\usepackage{subfigure}
\usepackage{color}
\usepackage{amsfonts,amssymb}
\usepackage[colorlinks,
            linkcolor=red, 
             anchorcolor=red, 
            citecolor=green, 
            ]{hyperref}

\usepackage{algorithm}
\usepackage{algorithmic}

\usepackage{array}
 \usepackage{mathrsfs}

\usepackage{amssymb}
\usepackage{amsmath}
\usepackage{cite}
\usepackage{url}
\usepackage{xcolor}
\usepackage{cite,graphicx,amsmath,amssymb}
\usepackage{subfigure}
\usepackage{fancyhdr}
\usepackage{mdwmath}
\usepackage{mdwtab}
\usepackage{caption}
\usepackage{amsthm}

\usepackage{mdwtab}
\usepackage{diagbox}

\usepackage{threeparttable}

\newtheorem{theorem}{Theorem}

\newtheorem{lemma}{Lemma}

\newtheorem{corollary}{Corollary}

\begin{document}
\title{When Full-Duplex Transmission Meets Intelligent Reflecting Surface: Opportunities and Challenges
\thanks{Manuscript received**, 2020; revised **, 2020; accepted **, 2020. The associate editor coordinating the review of this paper and approving it for publication was ***. }
\thanks{G. Pan, J. Ye, and M.-S. Alouni are with Computer, Electrical and Mathematical Sciences and Engineering Division, King Abdullah University of Science and Technology (KAUST), Thuwal 23955-6900, Saudi Arabia.}
\thanks{J. An is with the School of Information and Electronics Engineering, Beijing Institute of Technology, Beijing 100081, China. }
}

\author{Gaofeng Pan, Jia Ye, Jianping An, and Mohamed-Slim Alouini\vspace{-8mm}}


\maketitle

\begin{abstract}
Full-duplex (FD) transmission has already been regarded and developed as a promising method to improve the utilization efficiency of the limited spectrum resource, as transmitting and receiving are allowed to simultaneously occur on the same frequency band. Nowadays, benefiting from the recent development of intelligent reflecting surface (IRS), some unique electromagnetic (EM) functionalities, like wavefront shaping, focusing, anomalous reflection, absorption, frequency shifting, and nonreciprocity can be realized by soft-controlled elements at the IRS, showing the capability of reconfiguring the wireless propagation environment with no hardware cost and nearly zero energy consumption. To jointly exploit the virtues of both FD transmission and IRS, in this article we first introduce several EM functionalities of IRS that are profitable for FD transmission; then, some designs of FD-enabled IRS systems are proposed and discussed, followed by numerical results to demonstrate the obtained benefits. Finally, the challenges and open problems of realizing FD-enabled IRS systems are outlined and elaborated upon.

\end{abstract}



\section{Introduction}
Recent decades have witnessed the rapid proliferation of wireless devices, leading to an ever-increasing demand for seamless coverage, uninterrupted ubiquitous connectivity, and high data-rate transmission. However, due to the shortage of wireless spectrum, improving the spectrum efficiency has been regarded as an efficient way to fully exploit the limited spectrum resources that we currently possess, rather than seeking more spectrum resources towards to high-frequency bands such as millimeter waves and THz bands. As under full-duplex (FD) mode, the transceiver is capable of simultaneously transmitting and receiving on the same frequency band, FD transmission exhibits the potential to double the spectrum efficiency, bringing about the most efficient utilization of spectrum resources compared to half-duplex transmission \cite{Kim}.

On the contrary, nowadays, intelligent reflecting surface (IRS) is attracting more and more attention from both industry and academic research communities \cite{Wu}, because of its capability to reconfigure the wireless propagation environment via its soft-controlled functionalities of electromagnetic (EM) waves, e.g., wavefront shaping, focusing, anomalous reflection, absorption, frequency shifting, and nonreciprocity \cite{Abadal}. Specifically, differing from traditional radio frequency (RF) transceivers with modulation/demodulation, and RF modules, amplitude, frequency, and phase shafting are introduced by the numerous elements consisted in IRS through adjusting the coding matrices pre-stored in the field programmable gate array (FPGA) assembled with IRS while theoretically without any RF energy consumption \cite{Cui}. Consequently, the reconfigurability of the incident EM waves in amplitude, frequency, and direction domains is realized not only for reflection but also for transmission in three-dimensional space. 

Therefore, thinking about the case in which FD transmission is combined with IRS techniques, new degrees of freedom can be achieved to facilitate the design and construction of wireless communication systems to earn low-cost wireless coverage but high frequency efficiency. This is because no RF components are required at IRS, and IRS is then free of interference, leading to no necessity for costly self-interference cancellation. Some other benefits can also be attained from the EM functionalities of the IRS to improve/create the applications of FD transmission. For example, the reflect/transmission waves from the IRS can be adopted for some specific purposes, such as information delivery, cooperative jamming, wireless power transfer (WPT), artificial noise, and so on. So, generally, there are two directions to jointly implement FD transmission and IRS:

1) Instead of traditional RF relays, IRS can serve as a bridge to set up/enhance the FD transmission: i) co-time co-frequency FD transmission for line-of-sight (LOS) and non-LOS (NLOS) scenarios, and ii) frequency division FD for NLOS scenarios;

2) By utilizing the unique EM functionalities of IRS, the reflected/transmission waves of the incident ones from FD transmission links are employed for some other specific purposes, e.g., cooperative jamming, artificial noise, and WPT.

Motivated by these aforementioned observations, the main purpose of this article is to propose some designs for jointly exploiting the merits offered by FD transmission and IRS, which can serve as meaningful instances for future theoretical studies and practical engineering applications.

The rest of this article is organized as follows. In Section II, we present a brief introduction to the EM functionalities of IRS for FD transmission; in Section III, we elaborate on some designs of FD-enabled IRS systems, like FD-enabled IRS systems in LOS and NLOS scenarios, FD-enabled WPT, and FD-enabled secure communication. Some numerical results are presented to show the profits obtained from FD-enabled IRS systems in Section IV. The challenges and open problems of realizing FD-enabled IRS systems are discussed and summarized in Section V. Finally, the article is concluded in Section VI.

\begin{figure}[!htb]
\centering
\includegraphics[width= 3.2in,angle=0]{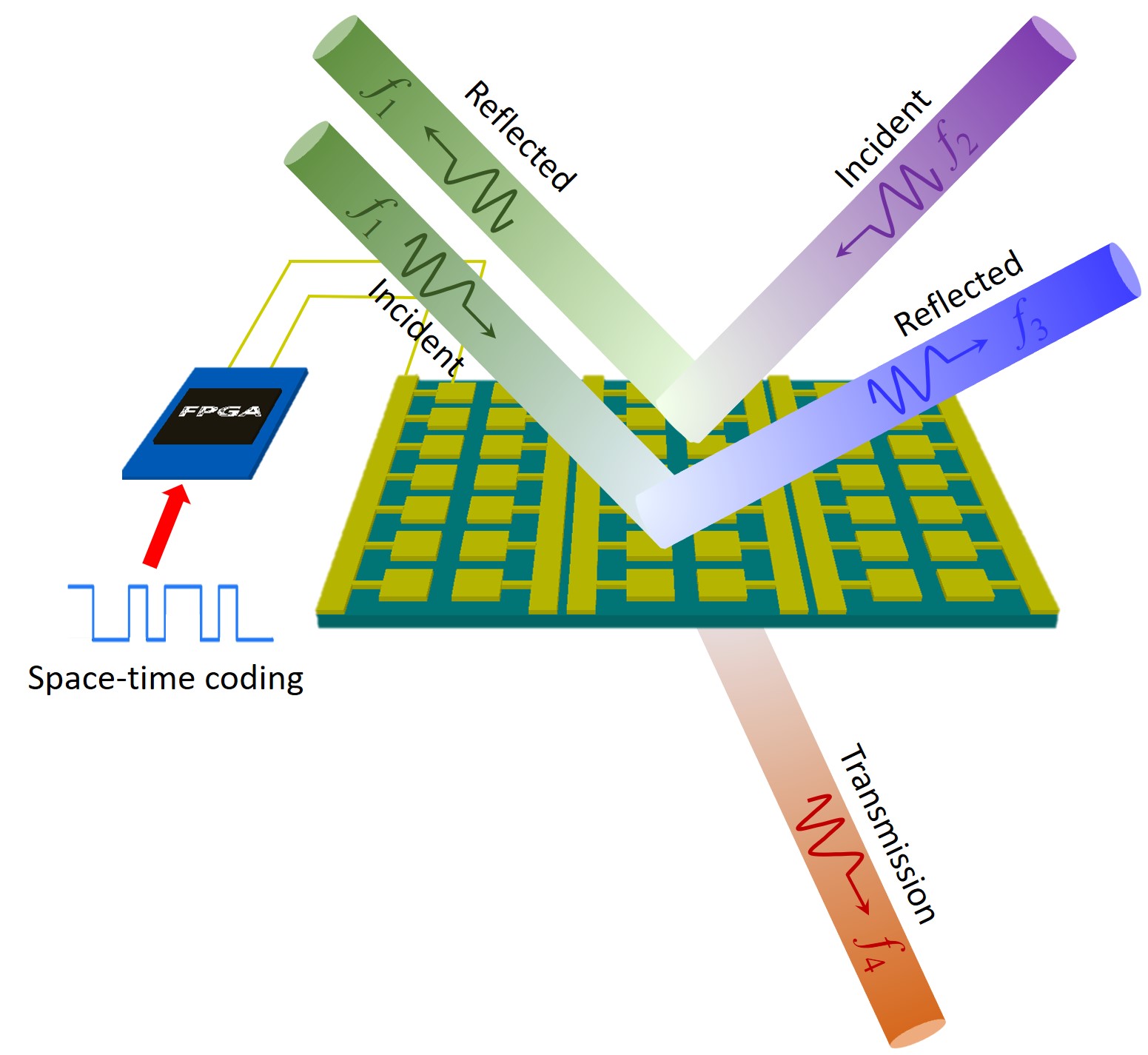}
\caption{EM functionalities of IRS.}
\label{fig_1}
\end{figure}

\begin{table*}[!htb]
\caption{EM Functionalities of IRS and Their Applications}
\label{T2}
\centering
\begin{center}
\begin{tabular}{|c|c|c|}
  \hline\hline
   EM Functionalities& Signal Domains & Applications\\ \hline
  Wavefront shaping& Amplitude& Cooperative jamming/WPT \\ \hline
  Focusing&  Amplitude&Multi-channel transmissions/cooperative jamming/WPT\\ \hline
  Anomalous reflection& Direction& Multi-user/directional/FD transmissions/WPT\\ \hline
  Absorption& Amplitude& Secure transmissions/interference cancellation\\ \hline
  Frequency shifting&Frequency& FD/Secure transmissions/WPT \\ \hline
  Nonreciprocity & Frequency \& Space& FD/Secure/Multi-channel transmissions/WPT \\
  \hline\hline
\end{tabular}
\end{center}
\end{table*}
\section{EM Functionalities of IRS for FD Transmission}
Generally, the electromagnetically reconfigurable elements in IRS are attributed to digital coding and programmable metamaterials. More specifically, as presented in \cite{Abadal}, the propagation of EM waves can be reconfigured, adjusted, and controlled via changing the coding matrices pre-stored in the FPGA assembled with IRS, leading to many EM functionalities such as wavefront shaping, focusing, anomalous reflection, absorption, frequency shifting, nonreciprocity, etc., as presented in Fig. \ref{fig_1}. Therefore, the possibilities of manipulating the propagating behaviors of EM waves readily open the door to new opportunities to reshape the implementation of wireless communications, e.g., FD transmissions focused upon in this article.

In the following, we will respectively introduce some EM functionalities of IRS enabling FD transmissions in various communication scenarios, as listed in Table \ref{T2}.

\subsection{Wavefront Shaping}
Multifunctional integrated coding metasurface can control both the transmitted and reflected wavefronts, by changing the polarization and direction of incident waves \cite{LZhang1}. In particular, the proposed metasurface can realize two functionalities simultaneously for a $45^{\rm{o}}$ polarized incident wave, which works in reflected and transmitted modes, respectively, while covering both sides of the metasurface. Moreover, as reported in \cite{RYWu}, the full-space wavefront shaping can be dynamically switched between reflection or transmission modes via adjusting the PIN diodes states.

Therefore, these aforementioned wavefront-shaping functionalities, like the switching function between reflection or transmission modes at the IRS elements and generating transmissions on the other side of the IRS, can be used to produce cooperative jamming signals to safeguard the information delivery.

\subsection{Focusing}
Programmable metasurfaces have been designed to focus EM fields \cite{XWan,KChen}, for example, a programmable Huygens¡¯ metasurface was designed to focus the transmitted fields \cite{KChen}. In other words, from the research in \cite{XWan} and \cite{KChen}, respectively, arose the possibilities of realizing multi-channel direct transmissions and controlling multiple and complex focal spots simultaneously at distinct spatial positions and reprogrammable in any desired fashion, with fast response time and high efficiency.

Then, thanks to these benefits, multi-channel and directional transmissions can be achieved, as well as cooperative jamming, via adjusting the transmission directions.

\subsection{Anomalous Reflection}
The reflection and scattering of THz waves can be tailored through configuring the digital coding elements of IRS, resulting in anomalous reflection and scattering behaviors \cite{LLiang}. This study also indicates that, through appropriate arranging the sequences of ¡°0¡± and ¡°1¡± elements of THz coding metamaterials, the far-field scattering patterns under a plane wave incidence vary from a single beam to two, three, and numerous beams, which depart, obviously, from the ordinary Snell's law of reflection \cite{LLiang}.

Accordingly, information eavesdropping can be improved by carefully designing the reflecting direction of EM waves to avoid potential eavesdroppers. Furthermore, cooperative jamming signals can also be generated via designing the number and directions of the reflecting beams to suppress the received signal-to-interference-plus-noise (SINR) at eavesdropping receivers.

\subsection{Absorption}
Absorption of EM waves has been studied and realized by using metamaterials in the past decade, due to their ability to decrease unwanted reflection. Passive metamaterial absorbers have been numerically designed to work in a fixed frequency band, while programmable metamaterials are considered to develop active absorbers profiting from their tunability with the change of external environment. Specifically, for instance, programmable metamaterials are utilized to absorb the incident waves at different frequencies by electrically controlling the biasing voltages on the varactors \cite{JZhao}, leading to extend working frequency range.

Intuitively, the absorption functionality of IRS can be applied to decrease the unnecessary and unintended reflection of the incident signals to reduce multi-path components and further degrade the received signal-to-noise (SNR) at malicious receivers for information protection purposes.

\subsection{Frequency Shifting}
The programmable metasurface is capable of shifting the frequency of reflecting waves from that of incident waves. For example, the frequency shifts of the time-reversed reflected wave can be programmed by the optimized space-time-coding matrix preloaded in the FPGA \cite{LZhang2}. In another instance, the reflection phase or amplitude can be modulated periodically with the predefined coding sequences by controlling the nonlinearity using time-domain digital-coding metasurfaces \cite{JZhao2}.

Frequency shifting can be exploited to generate an intended interfering signal to worsen the quality of the received signal at the eavesdroppers or to protect the target information transmission by changing the frequency.

\subsection{Nonreciprocity}
Nonreciprocity can be realized over the metasurface by inducing suitable spatiotemporal phase gradient to break reciprocity. A spatiotemporal phase gradient can be generated over the metasurface via suitably designing the space-time coding sequences, leading to the broken time-reversal symmetry and nonreciprocal wave reflections \cite{LZhang2}. Furthermore, in the same study simply switching between the states of nonreciprocity and reciprocity can be achieved by changing the codes stored in the FPGA.

Some potential applications can be designed to make use of the nonreciprocity induced by the programmable metasurface, e.g., safeguarding the information delivery by introducing interfering signals on an intended direction, and delivering energy to target terminals.

\section{The Designs of FD-Enabled IRS Systems}

\begin{figure}[!htb]
\centering
\includegraphics[width= 3.2in,angle=0]{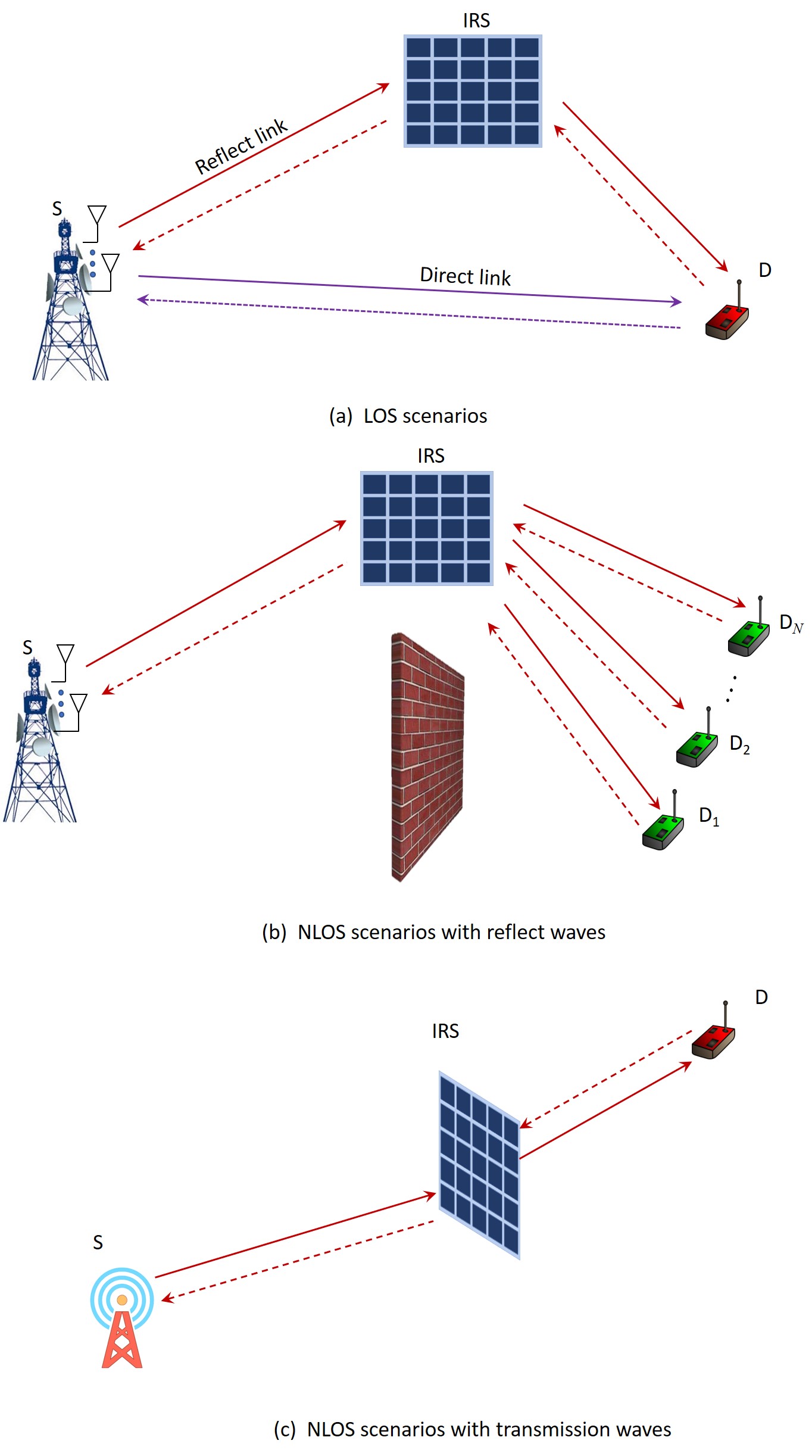}
\caption{FD Transmission in LOS \& NLOS Scenarios.}
\label{fig_2}
\end{figure}

In this section, several FD-enabled IRS applications will be introduced and illustrated to exploiting the unique EM functionalities of IRS.

Due to the flexibility arising from the numerous soft-controlled metasurfaces comprised in IRS, novel designing freedom is gained for wireless transmissions, like FD transmission, to satisfy the increasing demands of ubiquitous wireless connectivity. For example, the reflected properties of incident waves from different directions can be independently controlled without any interference \cite{LZhang1}, and Lorentz reciprocity can be broken by digital-controlled metasurface and reflected waves can be isolated in both space and frequency domains \cite{LZhang2}.

Generally, there are two types of applications for the IRS in FD-enabled IRS systems: 1) IRS serves as a component to build up or improve FD transmission links, which is similar to passive relay; and 2) IRS is employed to generate reflect/transmission waves of FD signals for other application purposes, e.g., WPT, cooperative jamming, and artificial noise.

Therefore, new opportunities are created for FD transmission in various scenarios, as depicted in the following subsections.

\subsection{FD Transmission in LOS \& NLOS Scenarios}
As suggested in Fig. \ref{fig_2}, FD transmission is available via the help of IRS by mainly making use of the functionalities of IRS, like incident waves from different directions can be independently reflected by separately controlling the elements of IRS via changing the codes in the FPGA \cite{LZhang1}.

\subsubsection{LOS Scenarios} Fig. \ref{fig_2}(a) illustrates a typical LOS scenario for FD transmission enabled IRS communication systems, in which a source (S) with multi-antenna communicates with a destination (D) under FD mode via the assistance of an IRS. Here, multiple copies of the transmitted from both S and D via both direct and reflect links can be obtained at D and S, respectively, resulting in diversity gains to improve the received SNR at D and S. Moreover, in this case, symmetric/asymmetric reflect functionality of IRS is utilized, as the location of S may not be symmetric to that of D with respect to IRS.

\subsubsection{NLOS Scenarios with Reflect Waves} Fig. \ref{fig_2}(b) presents a NLOS scenarios in which S communicates with multiple terminals (D$_i$, where $1 \le i \le N$ and $N \ge 1$) under FD mode. Under this case, the bi-direction communications between S and D$_i$ fully rely on the reflect links via the IRS. Obviously, there are multiple terminals communicating with S, the locations of which are randomly distributed in the blind zone of S's signal. However, though these terminals are out of the coverage of S because of the obstacle, FD transmission still can be reached by carefully designing the reflect matrix for the elements on the IRS.

\subsubsection{NLOS Scenarios with Transmission Waves} In Fig. \ref{fig_2}(c) shows another NLOS scenario to implement FD transmission, in which S and D are located at each side of an IRS and LOS transmission is not available. Similar to LOS scenarios, FD transmission can also be achieved in the absence of LOS signal for NLOS scenarios by making use of the functionality of IRS provided in \cite{LZhang1}.

\bigskip
Therefore, one thing here should be noted that four functionalities (wavefront shaping, focusing, anomalous reflection, and nonreciprocity) discussed in the last section are exploited in the scenarios shown in Fig. \ref{fig_2}. Also, as given in Fig. \ref{fig_2}(a) and (b), co-time co-frequency FD transmission is easily achieved for LOS and NLOS scenarios, and frequency division FD for NLOS scenarios, because IRS is capable of reflecting incident waves from different directions via independently controlling without any interference \cite{LZhang1}. Furthermore, FD transmission can be attained via frequency-division duplexing mode when frequency shifting is introduced by IRS to the reflected waves. In other words, separate transmission bands over the two transmission directions (namely, S$\rightarrow$IRS$\rightarrow$D/D$_i$ link and D/D$_i \rightarrow$IRS$\rightarrow$S link) are available by introducing frequency shifting at the IRS, shown in Fig. \ref{fig_2}.
\begin{figure}[!htb]
\centering
\includegraphics[width= 3.2in,angle=0]{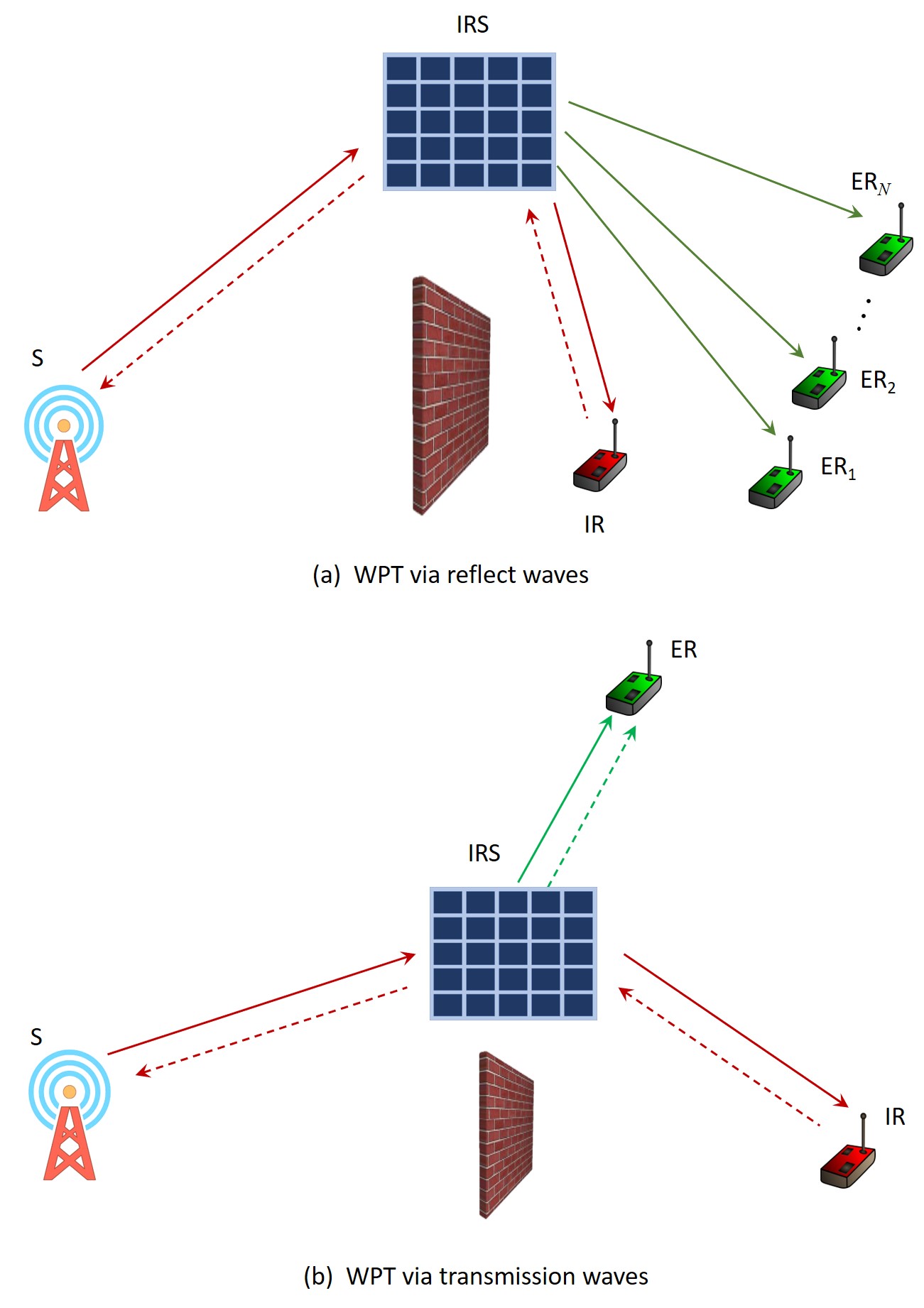}
\caption{FD-enabled WPT.}
\label{fig_3}
\end{figure}

\begin{figure*}[!htb]
\centering
\includegraphics[width= 7.in,angle=0]{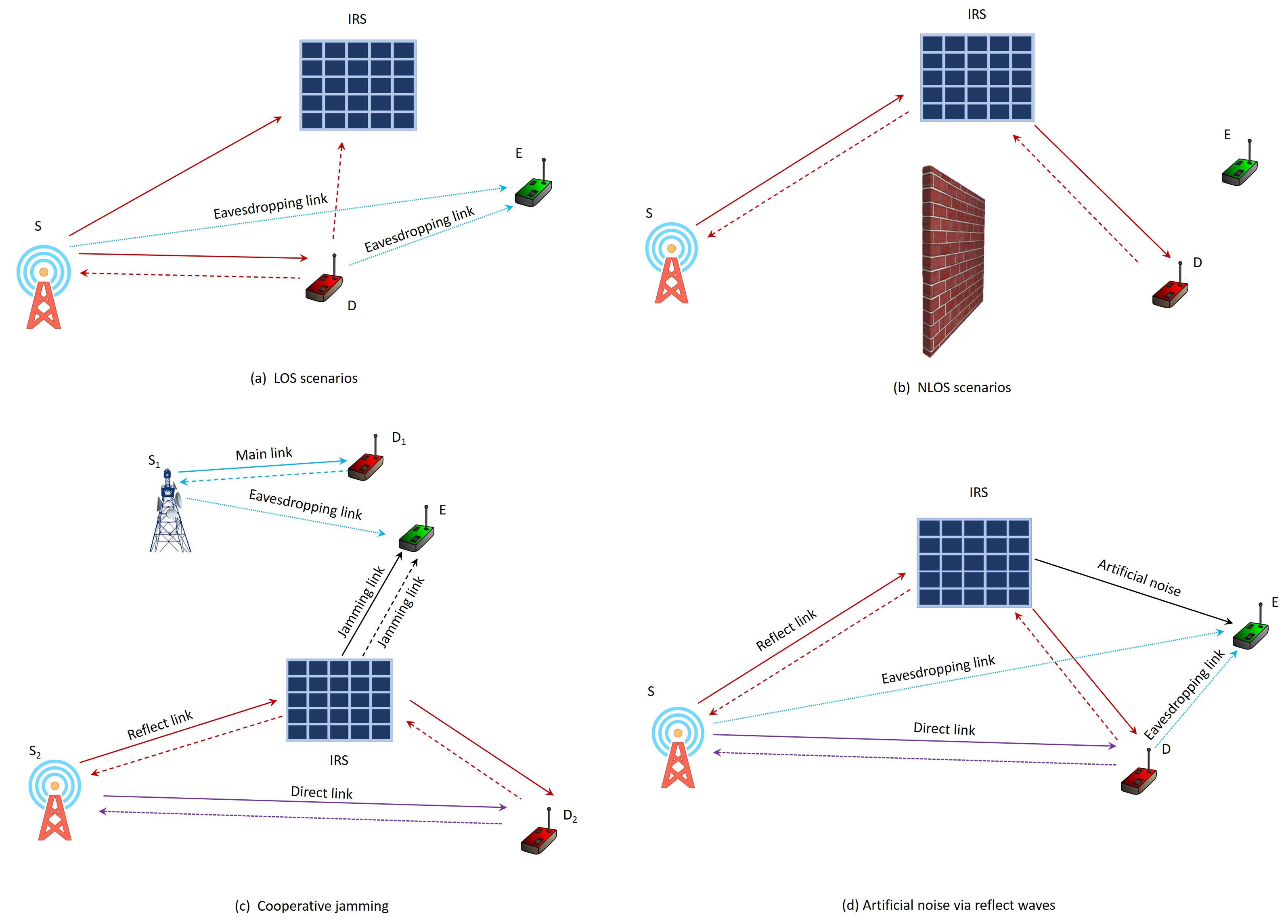}
\caption{FD-enabled Secure Communication.}
\label{fig_4}
\end{figure*}
\subsection{FD-enabled WPT}
In this subsection, we will introduce two cases of FD-enabled WPT scenarios: WPT via reflect waves and WPT via transmission waves, as depicted in Fig. \ref{fig_3}.

\subsubsection{WPT via Reflect Waves}
As shown in Fig. \ref{fig_3}(a), S and an information receiver (IR) communicate with each other under FD mode in NLOS scenarios, while there are multiple energy receivers (ER$_i$, where $1 \le i \le N$ and $N \ge 1$) harvesting energy from the transmitted signals from both S and IR and reflected by an IRS. So, it is easy to see that WPT can be realized by using part of reflected waves, and surely it can only happen on the premise that the quality of service (QoS) of the FD transmission between S and D is satisfied. Then, in such scenarios, all reflect waves can be fully exploited, either for bi-direction information delivery or for WPT purposes. Furthermore, the reflect matrix of the IRS can be optimized to achieve the best WPT performance while promising the FD transmission between S and D.

\subsubsection{WPT via Transmission Waves}
In this case, WPT is implemented on one side of the IRS while S and IR communicate with each other under FD mode on the other side of the IRS, shown in Fig. \ref{fig_3}(b). Clearly, the transmission waves of the transmitted signals from both S and IR are exploited to power ER. This is because the metasurface can work in reflected and transmitted mode simultaneously, covering both sides of the metasurface \cite{LZhang1}, which is the wavefront shaping functionality mentioned in the previous section. Similarly, there also exists an optimal matrix design problem at the IRS via allocating the signal components between the reflection for the FD information delivery and the transmission for WPT on the other side of the IRS, to realize the optimal WPT while promising the QoS of the FD information transmission.

\bigskip
Looking back to Fig. \ref{fig_3}, these two scenarios presented can be extended to LOS scenarios by considering the received signal via the direct links at the IR. The only difference is that the received SNR at the IR will be greatly improved, leaving more reflected/transmission waves for WPT in Fig. \ref{fig_3} (a) and (b), respectively.

\subsection{FD-Enabled Secure Communication}
As described in the previous section, the reflection/transmission behaviors of the incident waves can be intelligently controlled by IRS. Then, following physical layer security theory, there are some promising application scenarios for FD-enabled FD transmission to safeguard information security from the perspective of the physical layer.

\subsubsection{LOS Scenarios}
As shown in Fig. \ref{fig_4}(a), an eavesdropper (E) that is a little far from S, and D tries to overhear the transmitted information between S and D which work in FD mode, while an IRS is placed to improve the physical layer security of the FD transmission. Specifically, by unitizing the absorption functionality mentioned in the last section, the reflected waves of the transmitted signals from S and D can be absorbed by the IRS, then the received SNR of at E will be degraded without enough multi-path components. It should be noted that, since in LOS scenarios the QoS of the FD transmission between S and D can be easily guaranteed because of the strong signal over LOS links, IRS can be used to eliminate the multi-path components that will be contributed to the eavesdropping at E. On the other hand, it may fail when E is located close to S/D like D/S.

\subsubsection{NLOS Scenarios}
Fig. \ref{fig_4}(b) exhibits an NLOS scenario in which D and E are distributed out of the LOS  coverage of S, while an IRS is introduced to aid the communication between S and D, and E is trying to eavesdrop the FD information delivery between S and D. Differing from the case in Fig. \ref{fig_4}(a), the first task of the IRS adopted here is to bridge the bi-direction information transmission between S and D. Thanks to the unique functionalities provided by IRS, the physical layer security, in this case, is enhanced by adjusting the reflect directions of the incident waves transmitted by both S and D to avoid being eavesdropped by E.

\subsubsection{Cooperative Jamming}
Another application of IRS to improve physical layer security is to generate a cooperative jamming signal to degrade the received SINR at the eavesdropper, as depicted in Fig. \ref{fig_4}(c). In detail, E tries to overhear the information exchanged between S$_1$ and D$_1$ while S$_2$ communicates with D$_2$, and both of the two pairs work under FD mode. To protect the information delivery between S$_1$ and D$_1$, the transmission waves through the IRS from S$_2$ and D$_2$ can serve as cooperative jamming signal to induce the degradation of the eavesdropping SINR at E. Thus, it is another useful application of the transmission waves through the IRS, while the one in Fig. \ref{fig_2}(c) is for information delivery and the one in Fig. \ref{fig_3}(b) for WPT purpose.

\subsubsection{Artificial Noise via Reflect Waves}
One more scenario for FD-enabled secure communication is altering the application of the reflected waves to play as artificial noise to improve the physical layer security, as illustrated in Fig. \ref{fig_4}(d). One can see that S and D communicate with each other under FD mode, while an IRS serves a helper to enhance diversity gain and E attempts to overhear the information transmitted by both S and D. As the elements of the IRS can be independently and dynamically controlled to generate arbitrary focusing points \cite{XWan}, artificial noise signals can be sent to deteriorate the eavesdropping at E. On the other hand, IRS can generate various frequency and phase shifting to the reflect waves directed to E to increase the demodulation difficulties at E.

\bigskip
One more thing that should be elaborated for all aforementioned scenarios is that the applications of IRS prefer static/quasi-static networks because rapid changing of the locations of the terminals will require corresponding fast responses from the control unit of the IRS to accurately adjust the directions of the reflect/transmission waves to maintain stable FD communication or other application purposes (e.g., WPT and cooperative jamming).

\section{Numerical Results}
\begin{figure}[!htb]
\centering
\includegraphics[width= 3.in,angle=0]{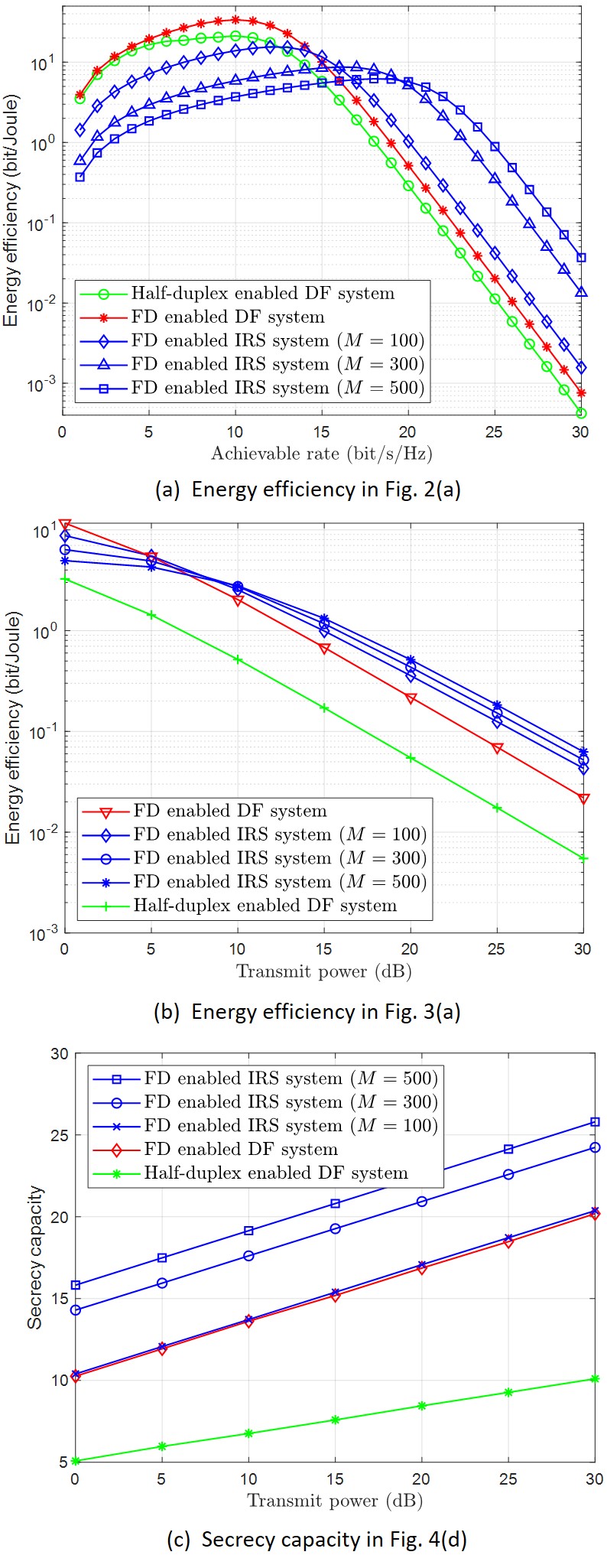}
\caption{Numerical results for FD-enabled IRS systems.}
\label{fig_5}
\end{figure}

In this section, some numerical results are given to show the performance of three designs of the proposed FD-enabled IRS systems, which are proposed in the previous section, over millimeter wave channels. The main parameters adopted here are set as: the transmit power at all devices is 100 mW while the one at the RF relay is 200 mW, the noise power at all devices is $-64$ dB, the efficiency of the transmit power amplifier is 2, the distance among each device is 20 meter. For comparison purposes, FD-enabled decode-and-forward (DF), and half-duplex (HD) enabled DF systems are considered. In WPT and secure communication scenarios, we set 70\% of the elements on the IRS to be adopted for energy transferring and artificial noise for simplification.

In Fig. \ref{fig_5}(a) and (b), energy efficiency (EE), which is defined as the number of the end-to-end transmitted bits per Joule, is taken into consideration. One can observe that, compared with FD-enabled DF and HD-enabled DF systems, the FD-enabled IRS system shows the best EE performance in large achievable rate and transmit power regions, respectively. As presented in Fig. \ref{fig_5}(c), the secrecy capacity of the FD-enabled IRS system outperforms the ones of the FD-enabled DF and HD-enabled DF systems.

All observations in Fig. \ref{fig_5} can be explained by the fact that these performance benefits are brought by the zero-RF-consumption elements on the IRS. Furthermore, the considered performance indices can be improved while the number of the elements ($M$) increases, profiting from the increased reflected waves from the IRS.

\section{Challenges and Open Problems}
Although numerical unique EM functionalities can be attained by IRS to facilitate the implementation of FD transmissions and extend the applications of FD transmission signals, some technical challenges and open problems require prompt solutions, which will be elaborated in the following.

\subsection{Network State Information Acquisition}
As discussed in Section III, to enable FD transmission via IRS, network state information including the locations of terminals and the number of terminals is required for designing the control matrix for the IRS. For example, as shown in Fig. \ref{fig_2}, accurate location information of S and D/D$_i$ is needed at the elements of the IRS to determine which reflect directions should be set to enable robust bi-directional transmissions between S and D. One more instance, the number of the ERs presented in Fig. \ref{fig_2}(b), $N$, must be provided for the controller of the IRS to allocate the elements of the IRS  for different purposes (namely, information delivery and WPT) to seek optimal network performance. Moreover, network state information is also quite meaningful for judiciously and optimally deploying IRS to facilitate and simplify the adjustment of the elements at the IRS in these such complicated application scenarios of FD-enabled IRS systems.

\subsection{Channel State Information (CSI) Acquisition}
As pointed out in \cite{Wu}, in general, the accurate CSI of the channels between the IRS and all terminals is necessary to realize various performance gains brought by IRS, which is a challenge for IRS, especially for passive IRS. Differing from traditional RF transceivers, there are no comparable calculation capacity and no RF signal transmitting and receiving capabilities at IRS, leading to the fact that traditional channel estimation schemes are normally not feasible in IRS systems. As accurate CSI is demanded to design, control, and optimize FD transmissions, novel channel estimation methods are highly yearned for FD-enabled IRS systems.

\subsection{Control Algorithm at the IRS}
Since IRS is made with numerous programmable elements, optimal algorithms with high efficiency are needed to be designed at the controller at the IRS to dynamically adjust the EM behaviors of these elements. For example, a general theory of space-time modulated digital coding metasurfaces should be proposed to achieve simultaneous manipulations of EM waves in both space and frequency domains, i.e., to control the propagation direction and harmonic power distribution simultaneously \cite{LZhang3}. Moreover, there exist numerical system factors, e.g., the application purposes of EM waves, various kinds of network topology, and different types of signal processing schemes, in the FD-enabled scenarios presented in Section III, inevitably affecting and increasing the difficulties that arise during the design of control algorithms, especially for large-scale deployment scenarios. Then, system-level performance analysis and optimization are essential to understand FD-enabled IRS systems, and then to achieve optimal system performance.

\subsection{Universality}
Obviously, the ideal IRS for various application purposes is to implement multiple functions within the same metasurface by exhibiting different behaviors simultaneously. As suggested by \cite{LZhang1}-\cite{QMa}, programmable metasurfaces with space-time reconfigurability are a promising way to implement software-driven EM control to obtain multiple desirable EM behaviors. Furthermore, as listed in Table I of \cite{Abadal}, the working frequency of most of the representative designs of coding metasurfaces ranges from 8 GHz to 2 THz, while one design listed there is designed for absorption functionality working in 5 GHz and another design in 2.75 GHz \cite{Kaina}. Thus, extending the operating frequency of IRS will promote IRS into practical applications and broaden the application scenarios.

\section{Conclusion}
In this article, we mainly presented some prototype application scenarios for FD-enabled IRS systems. First, we introduced the unique EM functionalities of IRS that facilitate the design and application of FD transmission. Second, we proposed and studied some designs of FD-enabled IRS systems in three aspects: FD-enabled transmission with IRS in LOS and NLOS scenarios, FD-enabled WPT with IRS, and FD-enabled secure communication with IRS. Moreover, we outlined and discussed the challenges and open problems of FD-enabled IRS systems in detail.

Based on the discussions and studies of the proposed designs of FD-enabled IRS systems, some useful conclusion and guidelines are obtained as follows:

\begin{itemize}
\item IRS can provide lost-cost and green assistance to set up/improve FD transmission links, especially for the devices out of the coverage of each other. Especially, co-time co-frequency FD transmission is easily achieved for LOS and NLOS scenarios, and frequency division FD for NLOS scenarios.

\item Thanks to IRS, other than information delivery, part/ all of the reflection and transmission waves of the incident waves over FD transmission links can be applied for more purposes, like cooperative jamming, artificial noise, and WPT.

\item The core technical challenge for realizing FD-enabled IRS systems is to develop the control algorithm for the controller of the IRS to accurately, optimally, and in a timely way adjust the EM functionalities of the elements at the IRS.

\item The accuracy of the network/channel sate information is the key factor affecting the efficiency and effectiveness of FD-enabled IRS systems.

\end{itemize}


\begin{thebibliography}{1}



\bibitem{Kim}
D. Kim, H. Lee, and D. Hong, ``A survey of in-band full-duplex transmission: From the perspective of PHY and MAC layers," \emph{IEEE Commun. Surveys Tuts.}, vol. 17, no. 4, pp. 2017-2046, Fourthquarter 2015.


\bibitem{Wu}
Q. Wu and R. Zhang, ``Towards smart and reconfigurable environment: Intelligent reflecting surface aided wireless network," \emph{IEEE Commun. Mag.}, vol. 58, no. 1, pp. 106-112, Jan. 2020.



\bibitem{Abadal}
S. Abadal, T. Cui, T. Low, and J. Georgiou, ``Programmable metamaterials for software-defined electromagnetic control: Circuits, systems, and architectures," \emph{IEEE Trans. Emerg. Sel. Topics Power Electron.}, vol. 10, no. 1, pp. 6-19, Mar. 2020.

\bibitem{Cui}
T. Cui, M. Qi, X. Wan, et al. ``Coding metamaterials, digital metamaterials and programmable metamaterials," \emph{Light Sci. Appl.}, vol. 3, e218, 2014.

\bibitem{LZhang1}
L. Zhang et al., ``Transmission-reflection-integrated multifunctional coding metasurface for full-space controls of electromagnetic waves,¡± \emph{Adv. Funct. Mater.}, vol. 28, no. 33, Jun. 2018, Art. no. 1802205.


\bibitem{RYWu}
R. Y. Wu et al., ``Digital metasurface with phase code and reflection¨C transmission amplitude code for flexible full-space electromagnetic manipulations,¡± \emph{Adv. Opt. Mater.}, vol. 7, no. 8, 2019, Art. no. 1801429.

\bibitem{XWan}
X. Wan et al., ``Multichannel direct transmissions of near-field information,¡± \emph{Light, Sci. Appl.}, vol. 8, no. 1, pp. 1¨C8, Jul. 2019.


\bibitem{KChen}
K. Chen et al., ``A reconfigurable active Huygens¡¯ metalens,¡± \emph{Adv. Mater.}, vol. 29, no. 17, Feb. 2017, Art. no. 1606422.


\bibitem{LLiang}
L. Liang et al., ``Anomalous terahertz reflection and scattering by flexible and conformal coding metamaterials,¡± \emph{Adv. Opt. Mater.}, vol. 3, no. 10, p. 1311, Jul. 2015.





\bibitem{JZhao}
J. Zhao, Q. Cheng, J. Chen, M. Q. Qi, W. X. Jiang, and T. J. Cui, ``A tunable metamaterial absorber using varactor diodes,¡± \emph{New J. Phys.}, vol. 15, no. 4, Apr. 2013, Art. no. 043049.

\bibitem{LZhang2}
L. Zhang et al., ``Breaking reciprocity with space-time-coding digital metasurfaces,¡± \emph{Adv. Mater.}, vol. 31, no. 41, 2019, Art. no. 1904069.

\bibitem{JZhao2}
J. Zhao et al., ``Programmable time-domain digital-coding metasurface for non-linear harmonic manipulation and new wireless communication systems,¡± \emph{Nat. Sci. Rev.}, vol. 6, no. 2, pp. 231¨C238, Nov. 2018.

\bibitem{LZhang3}
L. Zhang et al., ``Space-time-coding digital metasurfaces,¡± \emph{Nature Commun.}, vol. 9, Oct. 2018, Art. no. 4334.


\bibitem{QMa}
Q. Ma et al., ``Controllable and programmable nonreciprocity based on detachable digital coding metasurface,¡± \emph{Adv. Opt. Mater.}, vol. 7, no. 24, Oct. 2019, Art. no. 1901285.


\bibitem{Kaina}
N. Kaina, M. Dupr¨¦, G. Lerosey, and M. Fink, ``Shaping complex microwave fields in reverberating media with binary tunable metasurfaces,¡¯¡¯ \emph{Sci. Rep.}, vol. 4, no. 1, Oct. 2014, Art. no. 6693.

\end{thebibliography}
\end{document}